\documentclass[conference,a4paper]{APSIPA2021}
\usepackage{multirow, cite}
\usepackage{graphicx}
\usepackage{amsmath}
\usepackage[psamsfonts]{amssymb}
\usepackage{amsxtra}
\usepackage{threeparttable}
\usepackage{enumerate}
\usepackage{epstopdf, epsfig,color}

\begin{document}

\title{A PRIVACY-PRESERVING IMAGE RETRIEVAL SCHEME USING A CODEBOOK GENERATED FROM INDEPENDENT PLAIN-IMAGE DATASET}

\author{%
\authorblockN{%
Kenta Iida\authorrefmark{1} and
Hitoshi Kiya\authorrefmark{1}
}
\authorblockA{%
\authorrefmark{1}
Tokyo Metropolitan University, Tokyo, Japan \\
E-mail: iida-kenta2@ed.tmu.ac.jp, kiya@tmu.ac.jp}
}

\maketitle
\thispagestyle{empty}

\begin{abstract}
In this paper, we propose a privacy-preserving image-retrieval scheme using a codebook generated by using a plain-image dataset.
Encryption-then-compression (EtC) images, which were proposed for EtC systems, have been used in conventional privacy-preserving image-retrieval schemes, in which a codebook is generated from EtC images uploaded by image owners, and extended SIMPLE descriptors are then calculated as image descriptors by using the codebook.
In contrast, in the proposed scheme, a codebook is generated from a dataset independent of uploaded images.
The use of an independent dataset enables us not only to use a codebook that does not require recalculation but also to constantly provide a high retrieval accuracy.
In an experiment, the proposed scheme is demonstrated to maintain a high retrieval performance, even if codebooks are generated from a plain image dataset independent of image owners' encrypted images.
\end{abstract}

\section{Introduction}
\color{black}
\noindent With the rapid growth of cloud computing, outsourcing images to cloud storage services and sharing photos have greatly increased.
Generally, images are uploaded and stored in a compressed form to reduce the amount of data.
In addition, most images include sensitive information, such as personal data and copyright information \cite{Intro1, Intro2}.
However, cloud providers are not trusted in general, so there is the possibility of data leakage and unauthorized use in cloud environments.
Therefore, various image identification, retrieval, and processing schemes have been studied for untrusted cloud environments  \cite{EIR1,EIR2,EIR3,EIR4,EIR5,EIR6,EIR7,EIR8,EIR9,EIR10, EIR11,EIR12,EIR13, PPIR1, PPIR2, PPIR3, ID1}.

For the above reasons, privacy-preserving image-retrieval methods should satisfy generally three requirements: 1) protecting visual information on plain images, 2) having a high retrieval performance for encrypted images, and 3) being applicable to compressible encrypted images.
\color{black}
To satisfy  both requirements 1) and 3), encryption-then-compression (EtC) systems have been developed for privacy-preserving image retrieval \cite{EtC1, EtC2, EtC3, EtC4, EtC6}.
\color{black}
In this paper, we focus on a block scrambling-based image encryption method that was proposed for EtC systems, where images encrypted by the method are referred to as ``EtC images."

\color{black}
To retrieve EtC images stored in a database, a codebook is generated from these EtC images, and extended SIMPLE descriptors (E-SIMPLEs) of the stored EtC images and a query are calculated by using the codebook.
As a result, the retrieval performance of the methods depends on the EtC images uploaded by image owners. 
Accordingly, the retrieval performance of the conventional methods degrades if the uploaded EtC images are not enough for calculating the codebook. 
In addition, the conventional methods require recalculation to handle each EtC image dataset in the encrypted domain.  

Due to such a situation, in this paper, a novel privacy-preserving image retrieval scheme using a codebook generated by using a plain-image dataset independent of uploaded EtC images is investigated. 
Accordingly, codebooks can be prepared by using a suitable plain dataset, so the recalculation of codebooks is not required.
In the proposed method, EtC images can be used as compressible encrypted images as well for the conventional schemes.
In an experiment, the proposed scheme is demonstrated to maintain a high retrieval performance, even if codebooks are generated from a plain image dataset independent of image owners’ encrypted images.

\section{Related work}
\subsection{Privacy-preserving content-based image retrieval}
\color{black}
\noindent Privacy-preserving content-based image retrieval (CBIR) schemes have been proposed to carry out content-based image retrieval without visual information in cloud environments. 
\color{black}
The schemes are classified into two approaches in terms of how to obtain descriptors as below.

In the first approach, descriptors are extracted from plain images by an image owner \cite{EIR1,EIR2,EIR4,EIR6,EIR7,EIR8,EIR9}, and these descriptors are required to be encrypted by an image owner.
In addition, some secret information is shared between the image owner and users \cite{EIR2,EIR9}. 
\color{black}
In contrast, in the second approach, descriptors are directly extracted from encrypted images without decryption by cloud providers\cite{EIR3,EIR10,EIR11,EIR12,EIR13, PPIR1}, and the image owner has no secret information that should be shared with users. 
The proposed scheme corresponds to this approach.
Bag-of-visual words (BOVW) model-based \cite{EIR3,EIR10,EIR13, PPIR1,PPIR2,PPIR3} and JPEG-based \cite{EIR11,EIR12} schemes are also in this approach.
\color{black}

In conventional schemes for EtC images, to calculate image descriptors of EtC images stored in a database, a codebook is generated from the stored images.
Thus, cloud providers can not control retrieval performances because the codebook depends on the stored images.
However, the use of a  plain-image dataset independent of the uploaded EtC images has never been considered.
Therefore, the retrieval with such a dataset is considered in this paper.

\subsection{EtC images\label{sec:EtC}}
\begin{figure}[t!]
\includegraphics[width=85mm]{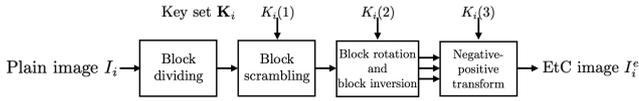}
\caption{Generation of EtC images\label{fig:enc}}
\end{figure}

\begin{figure}[t!]
\begin{center}
\begin{tabular}{c}
\begin{minipage}{0.5\hsize}
  \begin{center}
   \includegraphics[width=20mm]{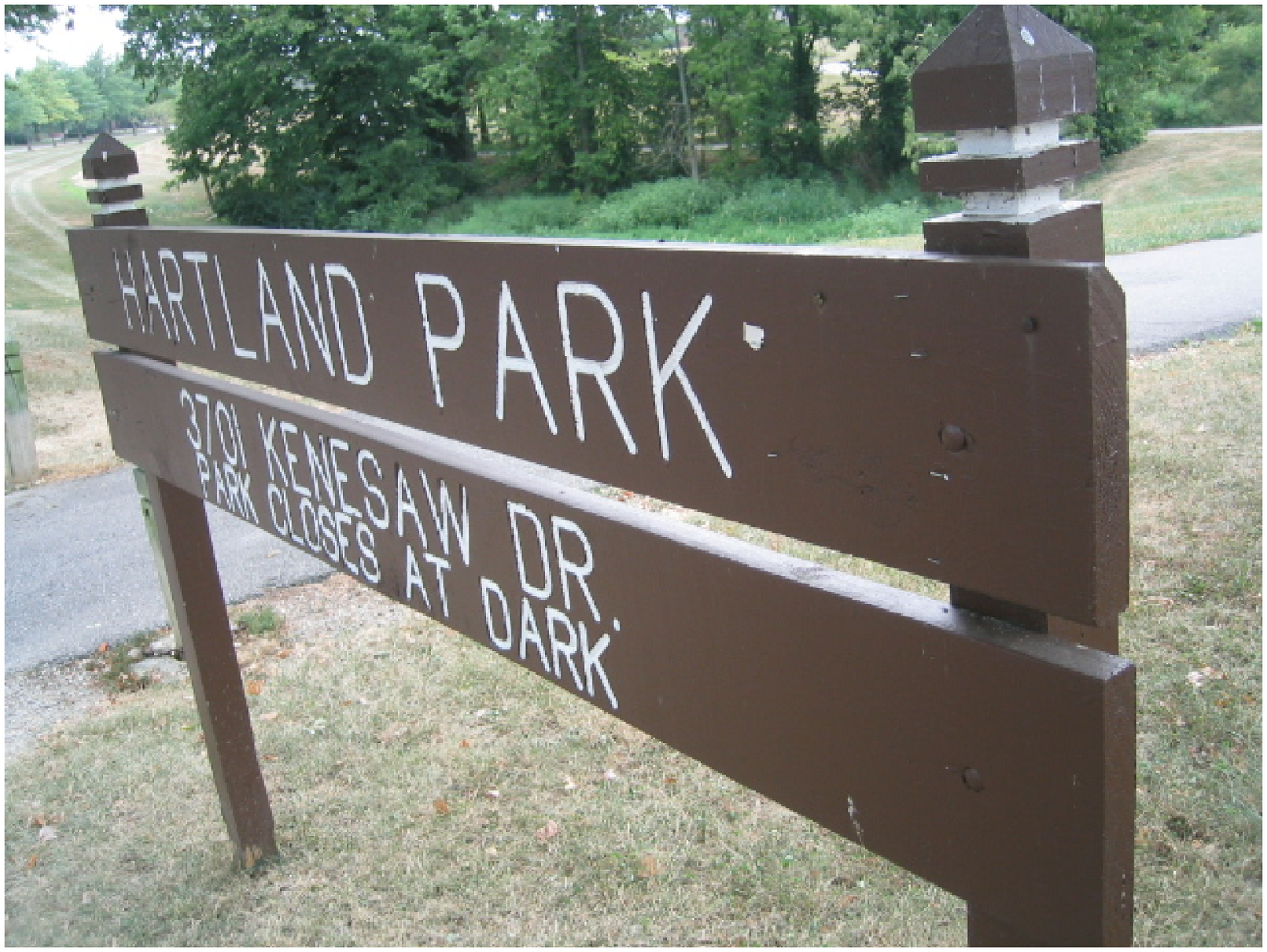}
     \hspace{20mm}(a) Plain image
  \end{center}
 \end{minipage}
\begin{minipage}{0.5\hsize}
  \begin{center}
   \includegraphics[width=20mm]{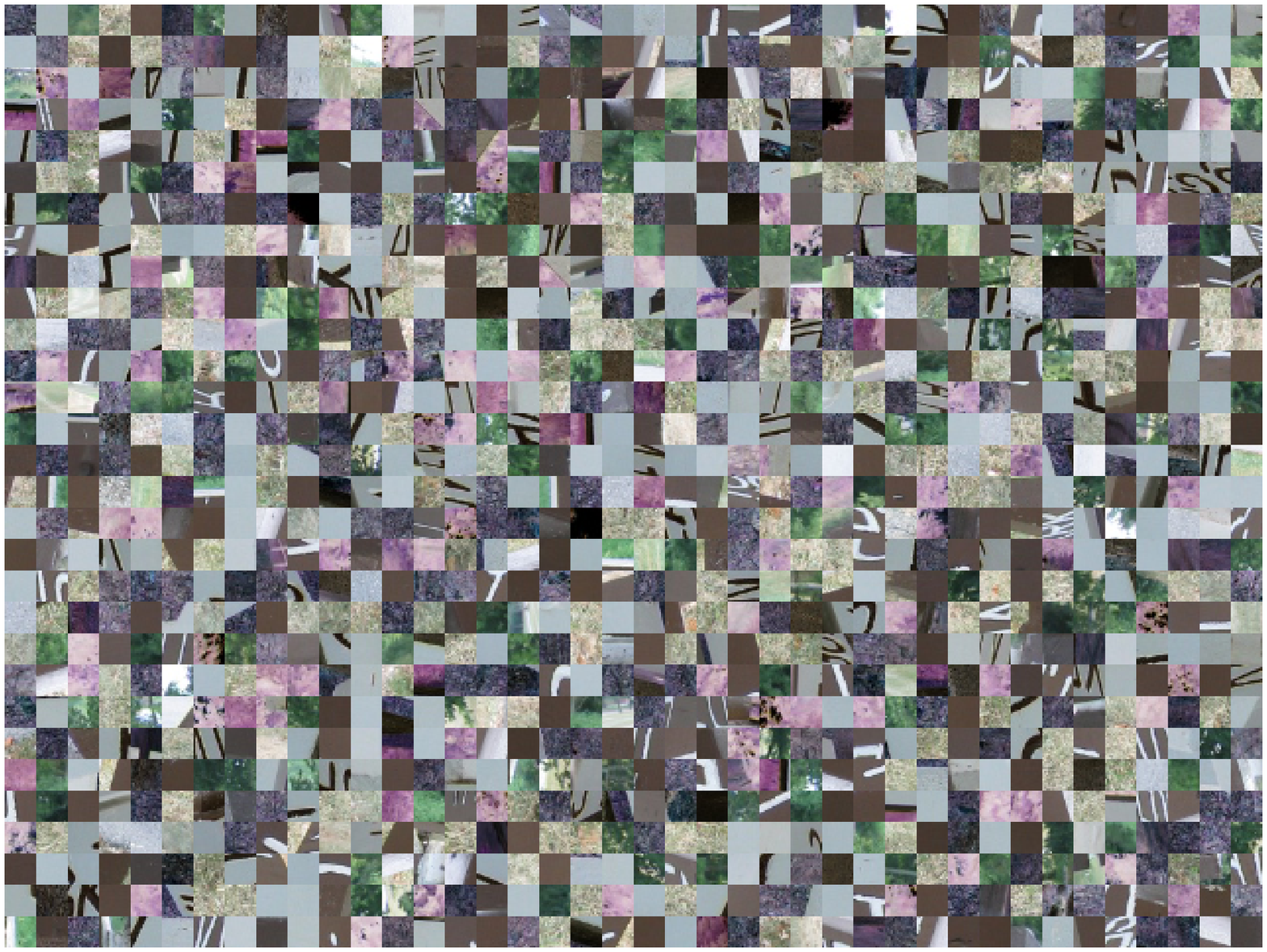}
     \hspace{20mm}(b) EtC image
  \end{center}
 \end{minipage}
\end{tabular}
\caption{Example of plain  and encrypted images\label{fig:encgen}}
 \end{center}
\end{figure}

\color{black}
\noindent  In this paper, we focus on EtC images, which are images encrypted by using a block-wise encryption method for EtC systems \cite{EtC2, EtC3, EtC4, EtC6}.
\color{black}
EtC images have almost the same compression performance as those of plain images but also enough robustness  against various ciphertext-only attacks including jigsaw puzzle solver attacks \cite{EtCattack1, EtCattack2, EtCattack3, EtCattack4, PPIR1}.
\color{black}
EtC images are generated by following the procedure as below (see Figs. \ref{fig:enc} and  \ref{fig:encgen}).
\color{black}

\begin{enumerate}[(a)]
\item
Divide image $I_i$ with $X \times Y$ pixels into non over-lapping $16 \times 16$ blocks. 
\item
Permute randomly $\lfloor \frac{X}{16} \rfloor \times \lfloor \frac{Y}{16} \rfloor$ divided blocks by using a random integer  secret key ${ K}_i(1)$.
\item
Rotate and invert randomly each divided block  by using a random integer  secret key ${ K}_i(2)$.
\item
Apply negative-positive transformation to each block by using a random binary integer generated by secret key $K_i(3)$ to obtain encrypted image $I_i^e$. In this step, a transformed pixel value in the $j$th block $B_j$, $p'$ is computed by
\begin{equation}
\label{eq:np}
\begin{cases}
p'=p,\ r(j)=0,\\
p' = 255 - p,\ r(j)=1,
\end{cases}
\end{equation}
where $r(j)$ is a random binary integer generated by $K_i(3)$ under the probability $P(r(j)) = 0.5$, and $p$ is the pixel value of a plain image with 8 bpp.
\end{enumerate}

In this paper, images encrypted by using these steps are referred to as ``EtC images.'' 
$K_i(1)$, $K_i(2)$, and $K_i(3)$ are stored as a key set, $\mathbf{K_i}=[K_i(1),K_i(2),K_i(3)]$.

\subsection{Extended SIMPLE descriptors}
\color{black}
\noindent Extended SIMPLE descriptors (E-SIMPLEs) were proposed for  image retrieval of EtC images, and high retrieval performances were reported in the conventional schemes \cite{PPIR1,PPIR2,PPIR3}.
In the proposed scheme, E-SIMPLEs are used as image descriptors, as well as conventional schemes.

E-SIMPLEs are designed for avoiding the influences of three encryption operations used for generating EtC images.
E-SIMPLEs are calculated from images as below.
\color{black}
\begin{itemize}
\item[a)] Divide each image into non-overlapping $16\times16$-blocks and use each $16\times16$-block as a patch, where $16\times16$ corresponds to the block size of EtC images.
\item[b)] Extract a modified color and edge directivity descriptor (mCEDD) as a patch descriptor from each patch.
\item[c)] Generate a codebook with a size of $M$ by using the patch descriptors extracted from all images.
\item[d)] Calculate a histogram vector of each image by using the codebook and patch descriptors extracted from the image.
\item[e)] Obtain extended SIMPLE descriptors by weighting the histogram vectors and $l_2$ normalization.
\end{itemize}
\color{black}
To generate the codebook in step c), $k$-means clustering is applied to the extracted patch descriptors, \color{black}where the size of the codebook corresponds to the number of classes in k-means clustering.
The center of each class is defined as a visual word, \color{black}and the set of visual words is stored as a codebook.
By using this codebook, a histogram vector of the frequencies of these visual words included in the image is calculated in step d). 

\color{black}
In step e), all histogram vectors are weighted in order to obtain extended SIMPLE descriptors. 
When $N$ histogram vectors are generated from $N$ images in step d), the $m$th component of the $n$th vectors $v_n(m)$, $0\leq m < M$, $0 \leq n < N$, is calculated as below.
\begin{equation}\label{eq:weight1}
v_n(m)=(1+log(v_n'(m))).
\end{equation}
where $v_n'(m)$ represents the $m$th  component of the $n$th histogram vector. 
After that, $l_2$ normalization is applied to every weighted histogram vector.
\color{black}


\section{Proposed scheme}
\begin{figure*}[t!]
\centering
\includegraphics[width=180mm]{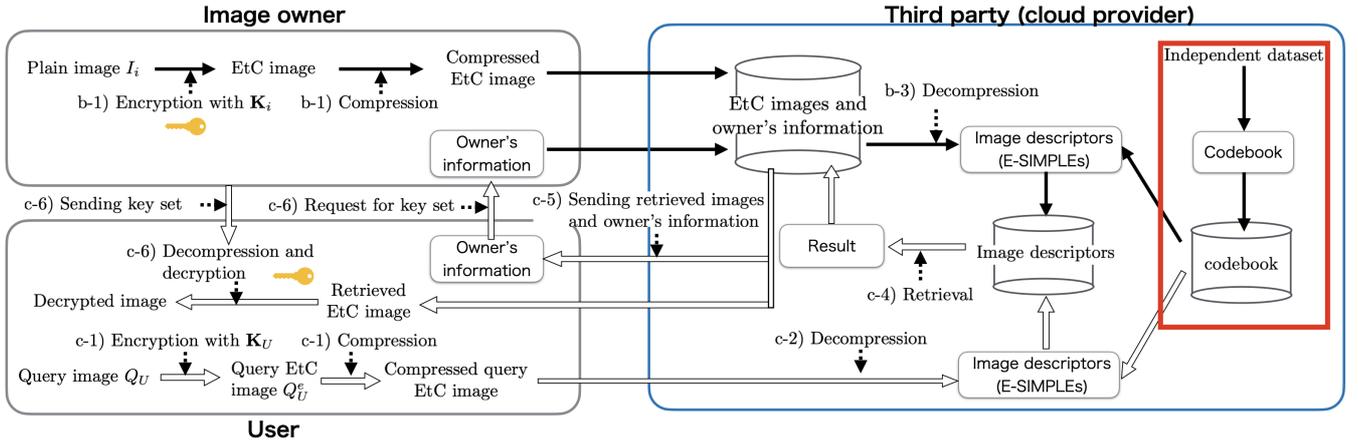}\\
\caption{System model of proposed scheme\label{fig:scenario}}
\end{figure*}
\subsection{System model}
\color{black}
\noindent  In the proposed scheme, the system model shown in Fig. \ref{fig:scenario} is used.
In this model, there are three roles: image owner, third party, and user, where the third party is not trusted.
\color{black}
The third party may do data leakage and unauthorized use of stored images, so the privacy-preserving of uploaded images is required.

In this model, the third party generates a codebook by using a plain-image dataset in advance, which is independent of EtC images uploaded by image owners.
An image owner uploads EtC images together with owner's information to the third party, and then the image descriptors are calculated with the pre-generated codebook.
When a user sends a query image to the third party, the third party retrieves images by using image descriptors of the query image and the stored descriptors.
The retrieved images are sent to the user with the owner's information,  and the user decrypts the EtC images with the key set received from the image owner. 

In this model, the third party has not only no visual information on plain images but also no key sets.

\subsection{Proposed image retrieval scheme\label{sec:proposed}}
\noindent Here, the details of each operation performed in the model  are summarized as below.

\subsection*{1) Generation process for codebook}
\noindent A third party prepares a codebook with a size of $M$ by using an independent training dataset as below.
\begin{enumerate}[(\textrm{a}-1)]
\item
Divide each image of the training dataset into non-overlapping $16\times16$-blocks, where $16\times16$ corresponds to the block size of EtC images.
\item
Extract a mCEDD as a patch descriptor from each extracted patch.
\item
Apply $k$-means clustering to the extracted patch descriptors under $M$ classes.
\item
Select each centroid vector as a visual word, and store the set of these visual words as a codebook.
\end{enumerate}

\subsection*{2) Generation process for image descriptors of uploaded images}
\noindent An image owner and  a third party perform the following processes to store an image and the image descriptors in the third party.

\begin{enumerate}[(\textrm{b}-1)]
\item
An image owner generates an EtC image from plain image $I_i$ with secret key set ${\bf K}_i$, and then compresses the EtC image with JPEG compression/ a lossless-compression method.
\item
The image owner uploads the compressed EtC images to a third party, and  the uploaded image is then stored in the third party.
\item
The third party obtains patches of the uploaded image by dividing the  image into non-overlapping $16\times16$-blocks after decompression.
\item
mCEDDs are extracted as patch descriptors of the uploaded image from the obtained patches.
\item
A histogram vector of the uploaded image is calculated from extracted patch descriptors by using the codebook generated in step (a-4).
\color{black}
\item
The third party calculates an E-SIMPLE by weighting  histogram vector in accordance with  Eq. \ref{eq:weight1} and applying $l_2$ normalization to the  histogram vectors.
\end{enumerate}

\subsection*{3) Retrieval process for query image} 
\noindent  To retrieve stored EtC images similar to a query image uploaded by a user, the following processes are carried out.

\begin{enumerate}[(\textrm{c}-1)]
\item
A user sends query image ${ Q}_U^e$ encrypted by using key set ${\bf K}_U$ to a third party after the compression, where ${\bf K}_U$ can be prepared by the user.
\item
The third party obtains patches from the query image after decompression, and extracts a mCEDD as a patch descriptor from each patch.
\item
The third party calculates an E-SIMPLE from the patch descriptors by using the codebook stored in the database.
\item
The third party computes the $l_2$ distance between every E-SIMPLE stored in the database and the E-SIMPLE of the query, and then retrieves similar images from the database.
\item
The retrieved images and the owner's information are returned to the user. 
\item
The user requests the data owner to send key set $\mathbf{K_i}$ for decrypting the EtC images received from the third party.
\end{enumerate}
\color{black}

\section{Experiment}
\begin{figure}[t]
\begin{center}
\begin{tabular}{c}
\begin{minipage}{0.23\hsize}
  \begin{center}
   \includegraphics[width=20mm]{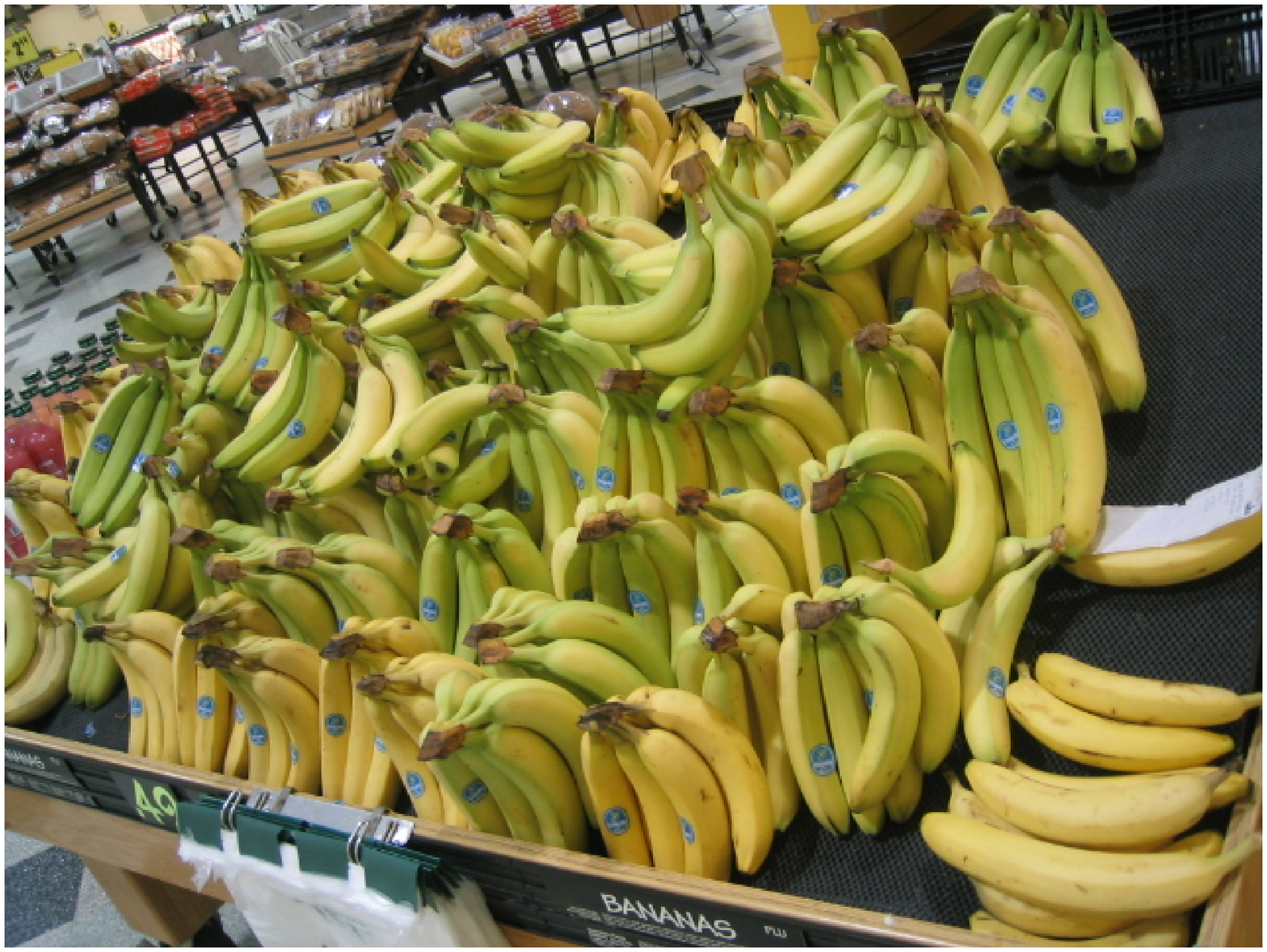}
  \end{center}
 \end{minipage}
\begin{minipage}{0.23\hsize}
  \begin{center}
   \includegraphics[width=20mm]{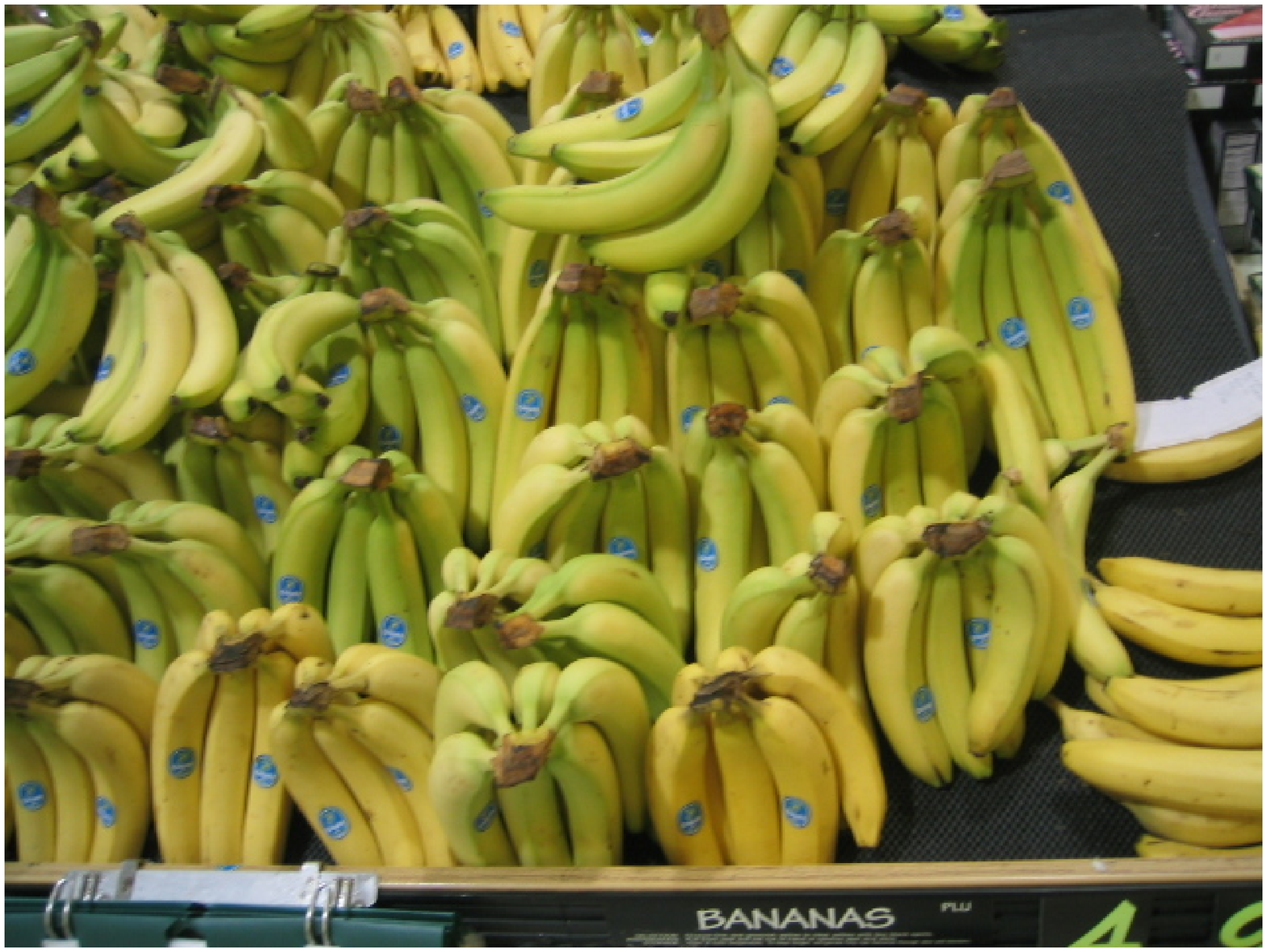}
  \end{center}
 \end{minipage}
 \begin{minipage}{0.23\hsize}
  \begin{center}
   \includegraphics[width=20mm]{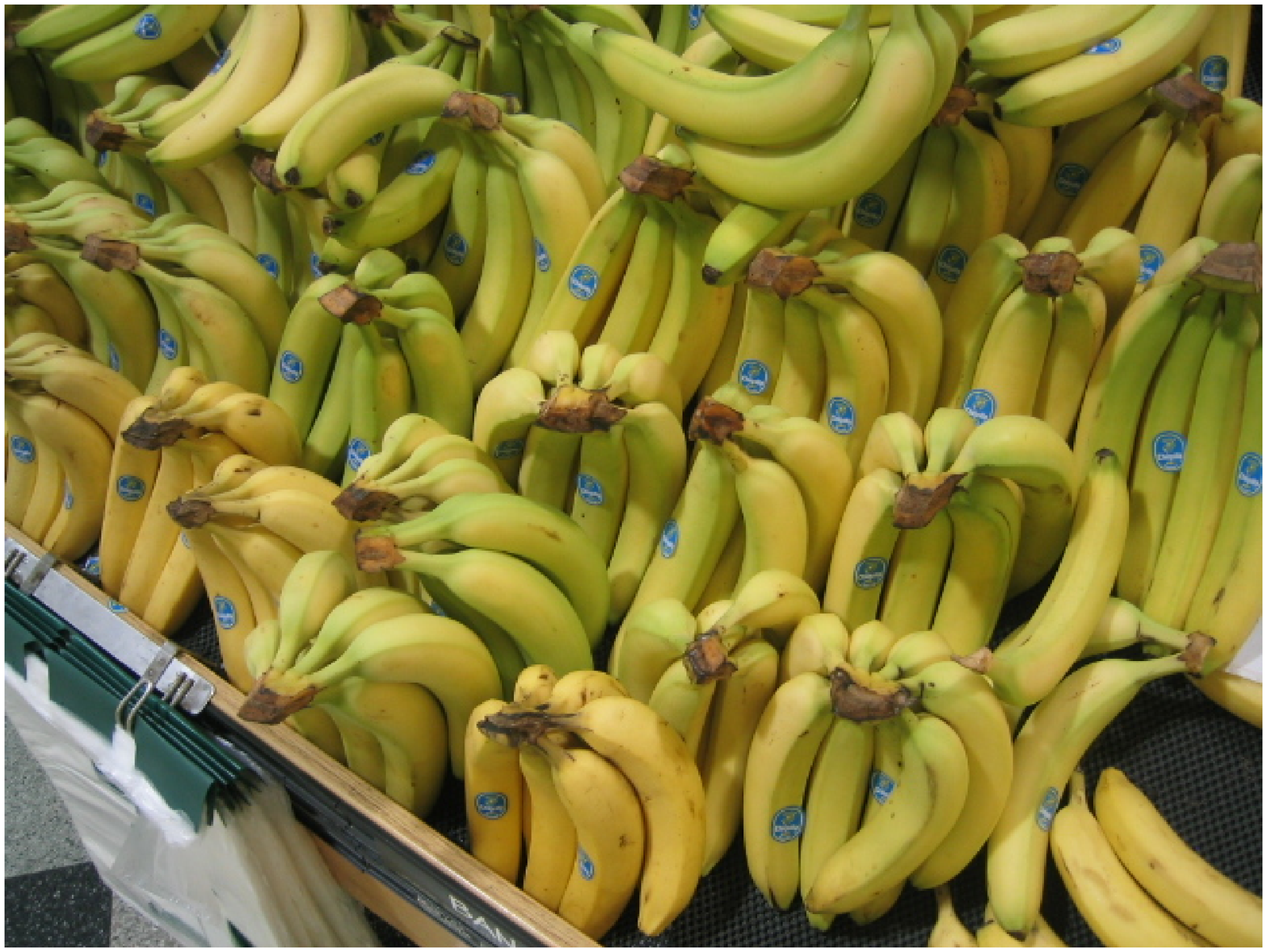}
  \end{center}
 \end{minipage}
 \begin{minipage}{0.23 \hsize}
  \begin{center}
   \includegraphics[width=20mm]{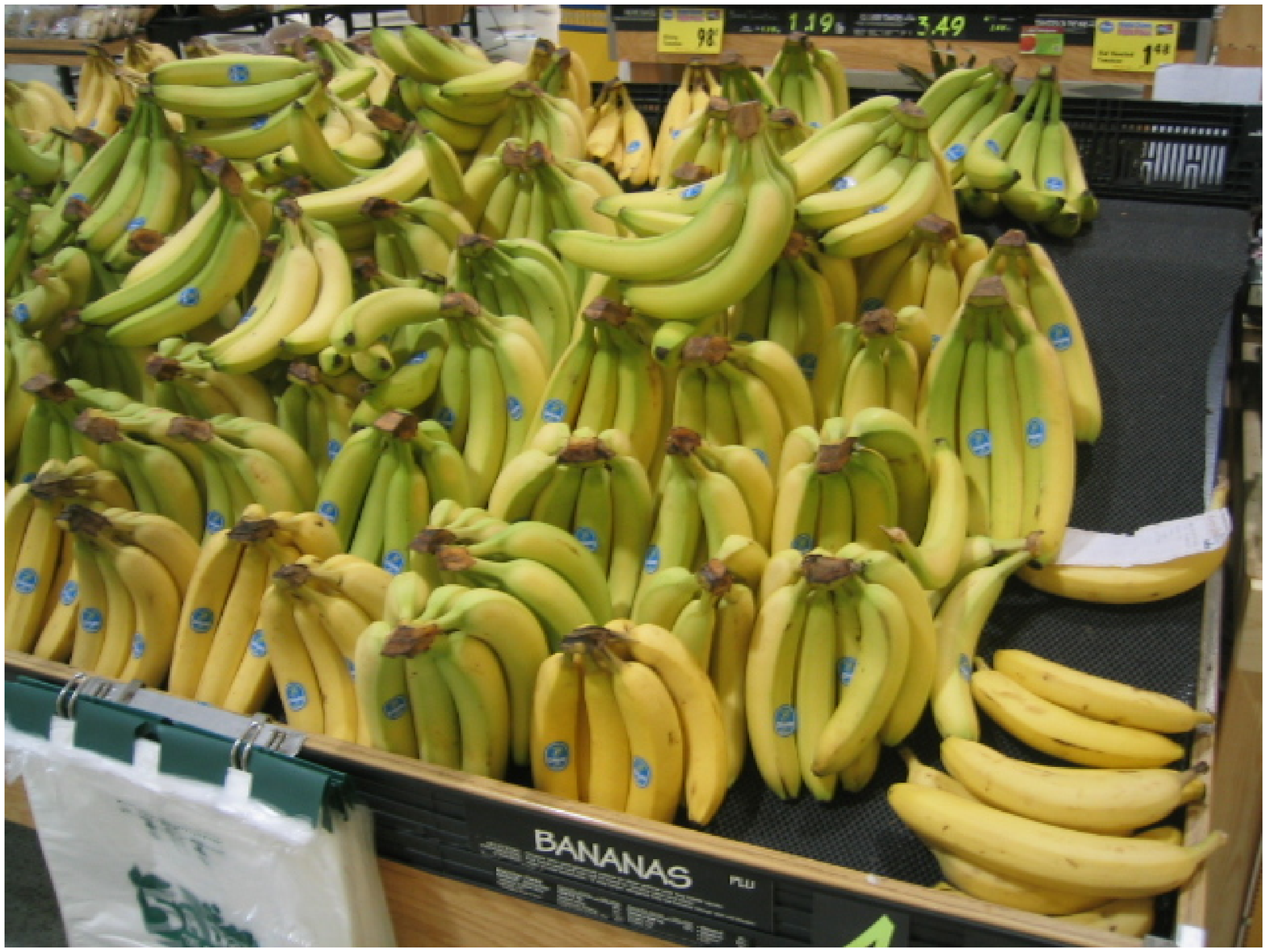}
  \end{center}
 \end{minipage}\\
 (a) Plain images \\
 
 \begin{minipage}{0.23\hsize}
  \begin{center}
   \includegraphics[width=20mm]{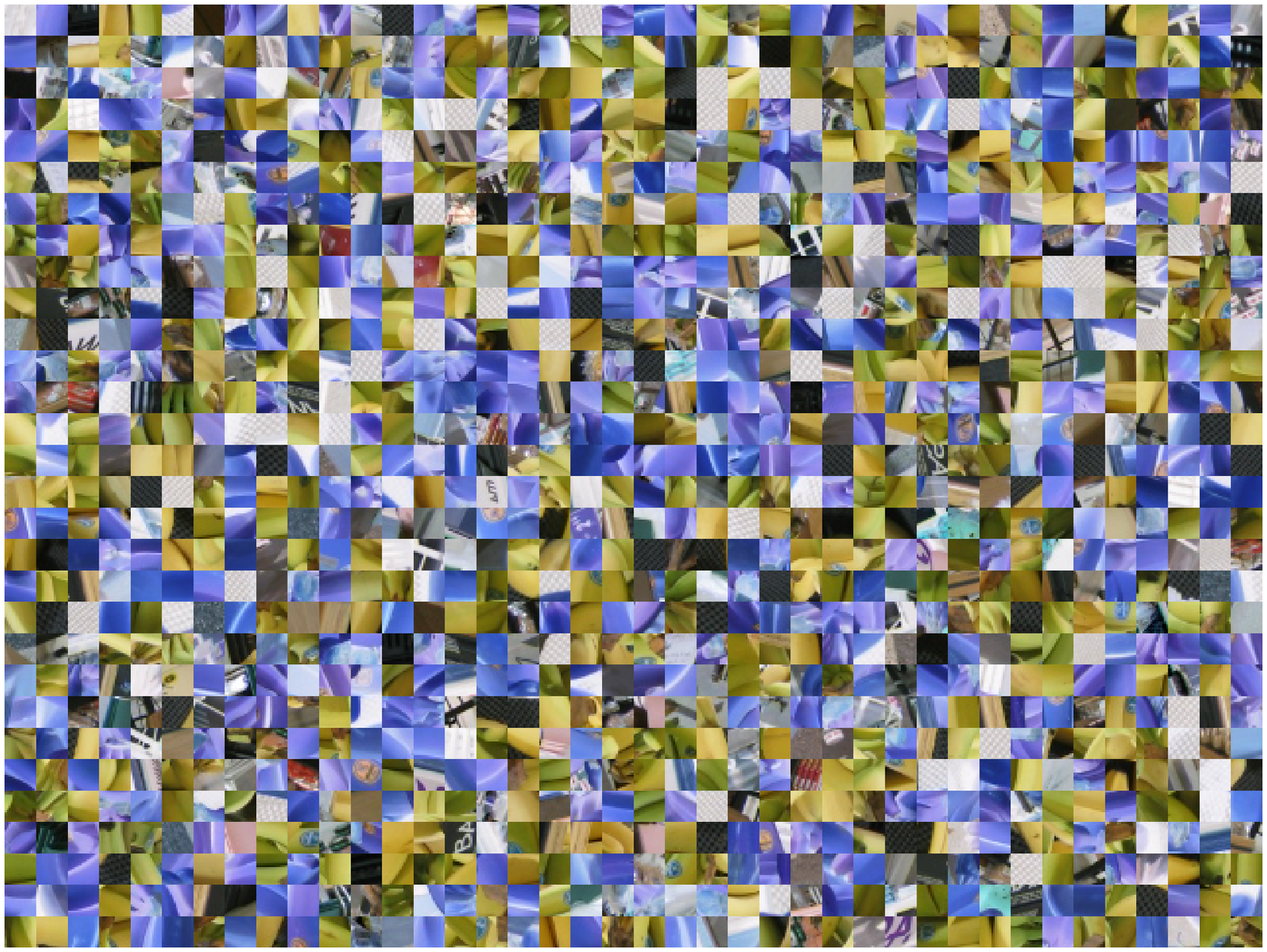}
  \end{center}
 \end{minipage}
\begin{minipage}{0.23\hsize}
  \begin{center}
   \includegraphics[width=20mm]{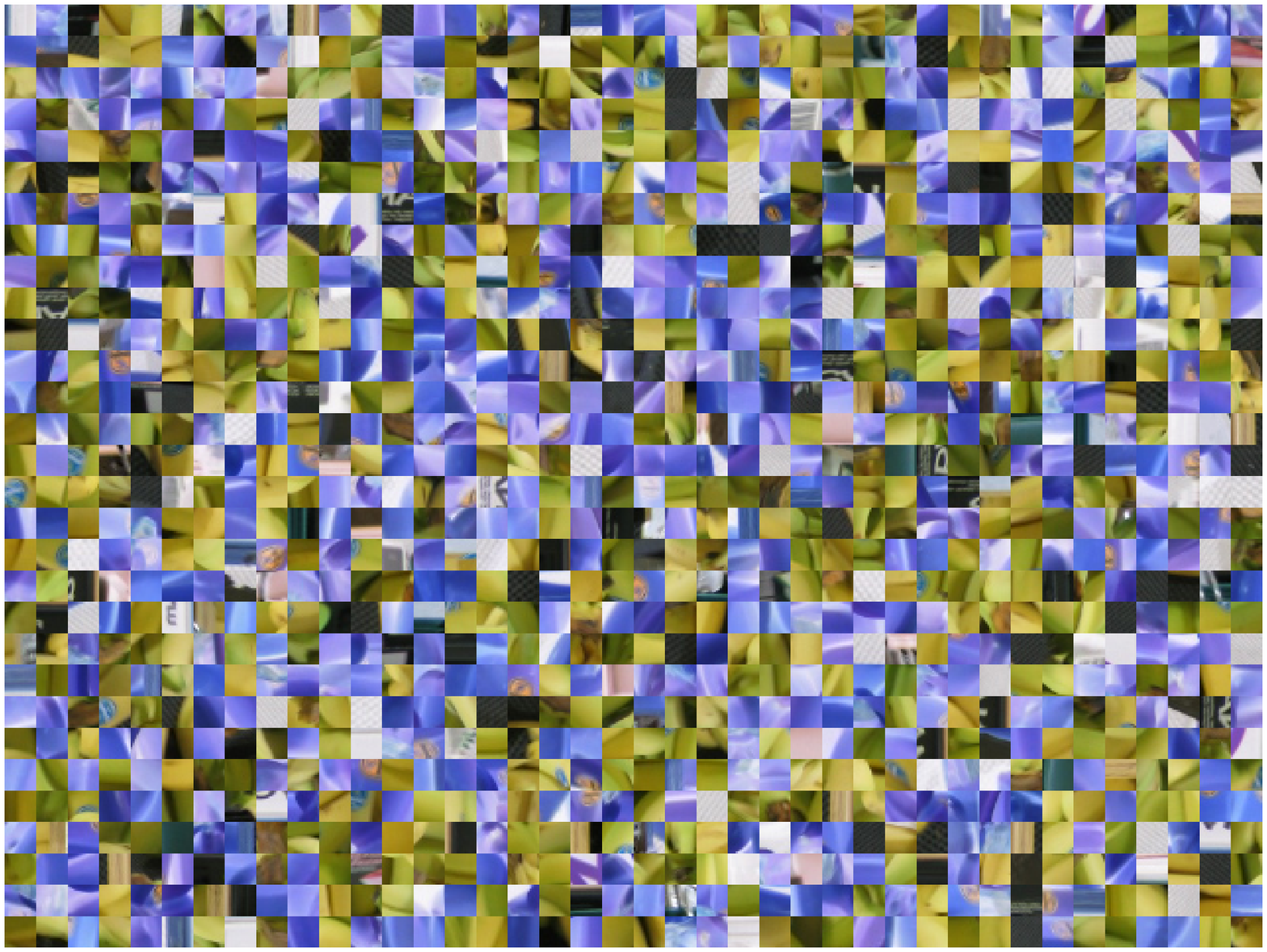}
  \end{center}
 \end{minipage}
 \begin{minipage}{0.23\hsize}
  \begin{center}
   \includegraphics[width=20mm]{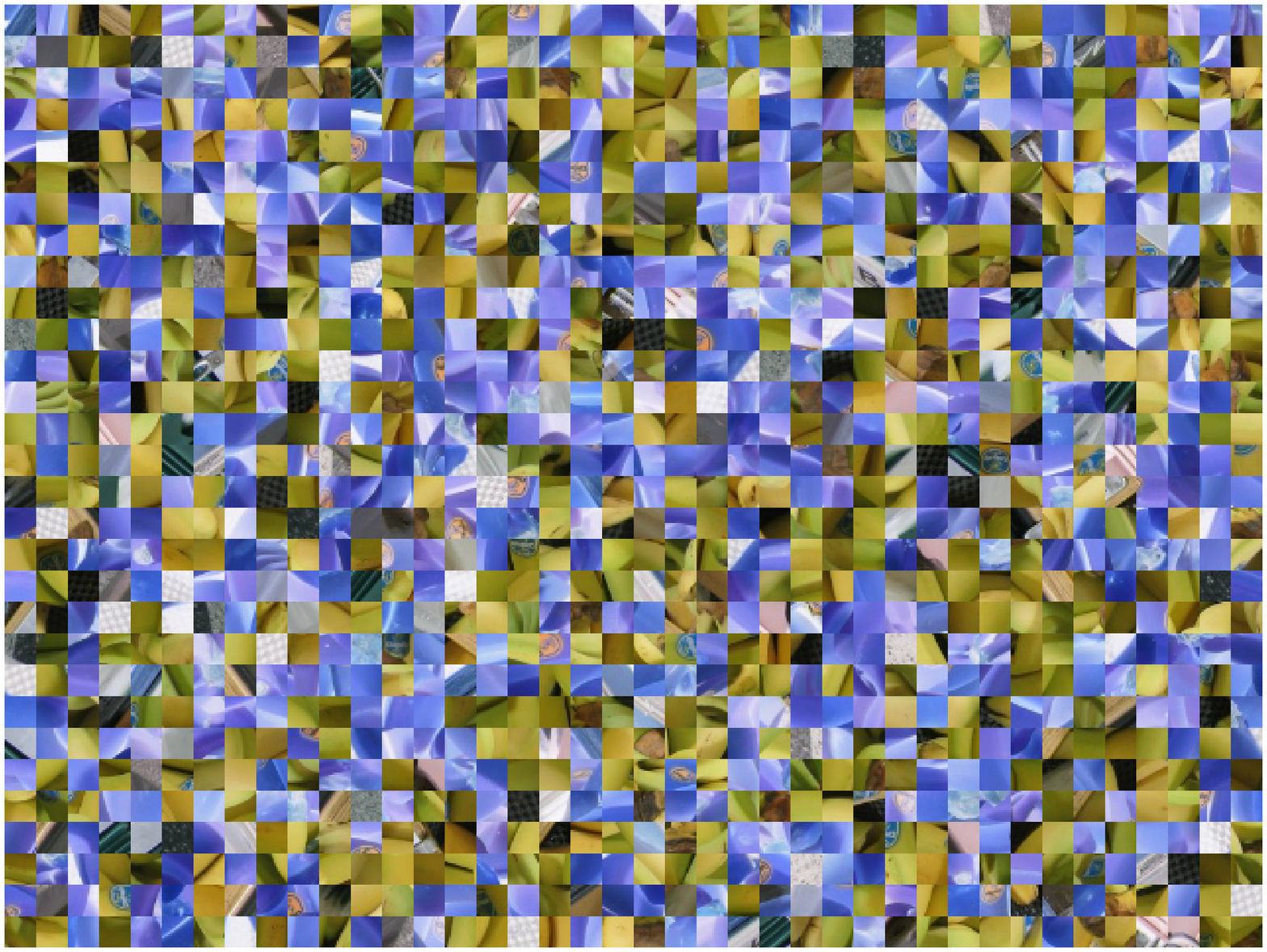}
  \end{center}
 \end{minipage}
 \begin{minipage}{0.23 \hsize}
  \begin{center}
   \includegraphics[width=20mm]{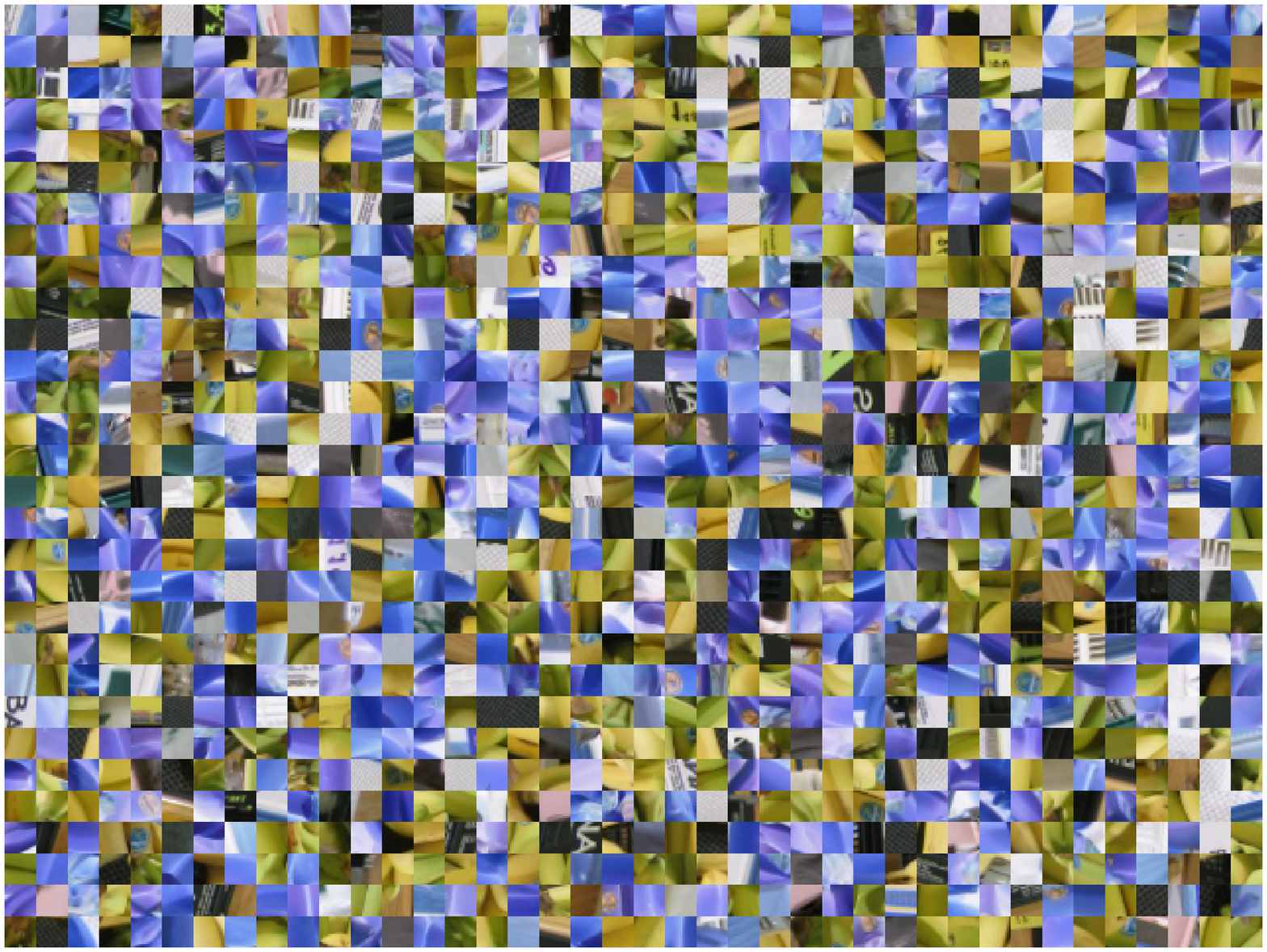}
  \end{center}
 \end{minipage}\\
(b) Corresponding EtC images \\
\end{tabular}
\caption{Image examples in group (UKbench dataset) \label{fig:exuk}}
 \end{center}
\end{figure} 

\subsection{Experiment setup}
\noindent  In this experiment, we used three datasets: the UKBench dataset\cite{uk}, the COREL  dataset\cite{corel}, and  the UCID dataset\cite{ucid}.
The COREL and the UCID datasets were used for generating codebooks.  1,000 images from No. 00000 to No. 00999 in the UKBench dataset were selected as query and user's images.
The 1,000 images in the UKBench dataset are classified into 250 groups and each group has four images (see Fig. \ref{fig:exuk}).
All the 1,000 images were used as images stored in the third party, and 250 query images were selected from the first image of each group.

\color{black}

The performance of the proposed scheme was evaluated in terms of mean average precision (mAP) scores in this experiment.
mAP scores are obtained from average precision values of $Q$ query images.
The average precision value of the $q$th query image is calculated as below, when  $G$ ground truth images of this query are contained  in $N$  images stored in a database.
\begin{equation}
AP_q=\frac{1 }{G} \sum_{n=1}^{N}\frac{TP@n}{n}\times f(n),
\end{equation}
where $TP@n$ represents the number of the true positive matches at the rank $n$ and $f(n)$ is defined by
\begin{equation}
f(n)=
\begin{cases}
1,\ \textrm{if the $n$th image is a ground truth image},\\
0,\ \textrm{otherwise.}
\end{cases}
\end{equation}
By using the average precision values for all $Q$ query images, mAP score is calculated as 
\begin{equation}
mAP=\frac{\sum_{q=0}^{Q-1} AP_q}{Q}.
\end{equation}

\
\begin{figure}[t!]
\includegraphics[width=87mm]{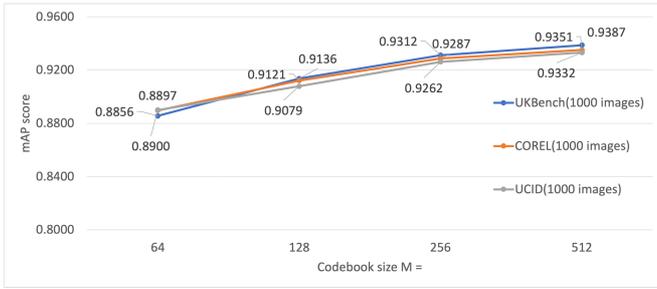}
\caption{Retrieval performance comparison under three codebooks generated from 1,000 images.\label{fig:resFull}}
\end{figure}

\begin{figure}[t!]
\includegraphics[width=87mm]{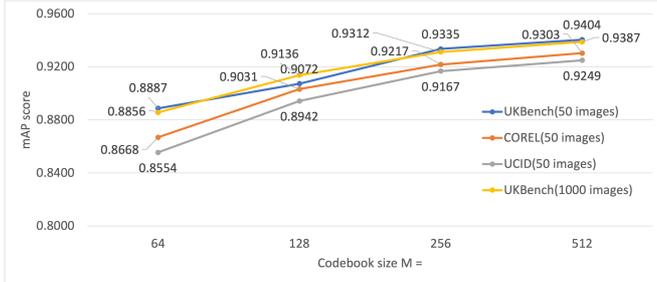}
\caption{Retrieval performance comparison under four codebooks generated from 50 or 1,000 images.\label{fig:resPart}}
\end{figure}


\begin{table}[t!]
\caption{Comparison with conventional CBIR methods using plain images.\label{tab:resUKorg}}
\centering
\begin{tabular}[b]{|c|c|c|c|c|}\hline
\multicolumn{2}{|c|}{Descriptor}&$M=$&mAP score\\\hline
\multicolumn{2}{|c|}{SCD \cite{scd} (plain)}& - & 0.9179\\\hline
\multicolumn{2}{|c|}{CEDD \cite{cedd} (plain)}& - & 0.8806\\\hline
\multicolumn{2}{|c|}{SURF \cite{surf}}&256&0.8304 \\ \cline{3-4}
\multicolumn{2}{|c|}{(plain)}&512&0.8355\\\hline
\multicolumn{2}{|c|}{Weighted SIMPLE with CEDD }&256&0.9300\\\cline{3-4}
\multicolumn{2}{|c|}{and random sampling (plain) \cite{SIMPLE}}&512&0.9481\\\hline
\multicolumn{2}{|c|}{Weighted SIMPLE with CEDD}&256&0.9000\\\cline{3-4}
\multicolumn{2}{|c|}{and SURF detector (plain) \cite{SIMPLE}}&512&0.9222\\\hline \hline
\multicolumn{2}{|c|}{Proposed scheme }&256&0.9287\\\cline{3-4}
\multicolumn{2}{|c|}{with COREL (EtC)}&512&0.9351\\\hline
\multicolumn{2}{|c|}{Proposed scheme  }&256&0.9262\\\cline{3-4}
\multicolumn{2}{|c|}{ with UCID (EtC)}&512&0.9332\\\hline
\end{tabular}
\end{table}
\color{black}
\subsection{Performance evaluation of  proposed scheme}
\noindent In Fig. \ref{fig:resFull}, retrieval performances under three codebooks generated from 1,000 images in each dataset are shown.
It was confirmed that mAP scores under the use of codebooks generated from images in the COREL and the UCID datasets were almost the same as those with codebooks generated from the uploaded images (UKBench), where the COREL and UCID datasets are independent of the UKBench dataset used by the image owner. 
From this figure, the proposed scheme was demonstrated to maintain a high retrieval performance even when the codebook was calculated by using a plain-image dataset independent of the uploaded dataset.

Next, retrieval performances under the use of codebooks generated from a smaller number of images were evaluated as shown in Fig. \ref{fig:resPart}. 
Compared with Fig. \ref{fig:resFull}, mAP scores for the COREL and UCID datasets degraded, but those for the UKBench dataset were maintained. 
From these results, when the codebook calculated from a small number of images was applied to images in other datasets, the reuse of the codebook provided a degraded performance. 
In contrast, the use of an independent dataset containing the enough number of images maintained a high performance. 
Accordingly, the proposed scheme that uses an independent dataset containing the enough number of images can maintain a high retrieval performance without the recalculation of codebooks.

\subsection{Comparison with conventional image retrieval scheme}
\noindent To compare the proposed scheme with conventional CBIR schemes that do not consider privacy protection, five image descriptors: scalable color descriptor (SCD) \cite{scd}, color and edge directivity descriptor (CEDD) \cite{cedd}, SURF \cite{surf}, weighted SIMPLE descriptor with random sampling\cite{SIMPLE}, and weighted SIMPLE descriptor with SURF detector\cite{SIMPLE}. 
From Table \ref{tab:resUKorg}, the proposed scheme was verified to have almost the same performance as those of the conventional ones, even when codebooks were generated from an independent dataset.

\section{Conclusion}
\noindent  A novel CBIR scheme using codebooks generated from a plain-image dataset independent of uploaded images was proposed for privacy-preserving image retrieval. In the proposed scheme,
EtC images can be used to reduce the amount of data, and extended SIMPLE descriptors are applied to avoid the influence of image encryption steps. 
In addition, codebooks are generated from a suitable plain-image dataset. Experimental results showed that the proposed scheme achieved a reasonable high retrieval performance, even when codebooks were generated from a plain image dataset independent of uploaded images.

\section*{Acknowledgment}
\noindent This study was partially supported by JSPS KAKENHI (Grant Number JP21H01327) and Support Center for Advanced Telecommunications Technology Research, Foundation (SCAT).

\bibliographystyle{IEEEbib}
\bibliography{ref}
\end{document}